%% file: paper.tex
\documentclass{llncs}
\usepackage{makeidx}  
\usepackage{booktabs} 
\usepackage{xspace}
\usepackage{caption}
\usepackage{listings}
\usepackage{lipsum}
\usepackage{courier}
\usepackage{array,multirow}
\usepackage{float}
\usepackage{graphicx}
\usepackage[perpage,para]{footmisc}
\usepackage{hyperref}
\usepackage{listings}
\usepackage[ruled,linesnumbered]{algorithm2e}
\usepackage{url}
\usepackage{subcaption}

\usepackage{tikz}
\usetikzlibrary{arrows,shapes.geometric,positioning,shapes.multipart,matrix,calc,patterns}
    \usetikzlibrary{topaths}
\usetikzlibrary{decorations.text}
\usepackage{pgfplots}
\usepackage{color}
\usepackage{pgfplotstable}

\lstdefinestyle{rm}{mathescape,basicstyle=\small\ttfamily}

\newcommand{\REM}[1]{}

\newcommand{\GPU}{\texttt{GPU}\xspace}
\newcommand{\CPU}{\texttt{CPU}\xspace}

\begin{document}
\title{Custom Code Generation for a Graph DSL}
 \author{Bikash Gogoi \inst{1}, Unnikrishnan Cheramangalath \inst{2}, Rupesh Nasre \inst{3}}%
 \institute{Department of CSE, Indian Institute of Technology, Madras, \email{bikashgogoi001@gmail.com}  \and
 Singapore University of Technology and Design, Singapore, \email{unnikrishnan\_cheramangalath@sutd.edu.sg} \and
 Department of CSE, Indian Institute of Technology, Madras, \email{rupesh@iitm.ac.in}  
}

\maketitle

\input{abstract}

\input{introduction}
\input{background}

\input{approach}

\input{results}

\input{related}
\input{conclusion}


\bibliographystyle{splncs03}
\bibliography{ref} 

\end{document}

%% file: abstract.tex
\begin{abstract}
We present challenges faced in making a domain-specific language (DSL) for graph algorithms adapt to varying requirements to generate a spectrum of efficient parallel codes. Graph algorithms are at the heart of several applications, and achieving high performance with them has become critical due to the tremendous growth of irregular data. However, irregular algorithms are quite challenging to parallelize automatically, due to access patterns influenced by the input graph -- which is unavailable until execution. Former research has addressed this issue by designing DSLs for graph algorithms, which restrict generality but allow efficient code-generation for various backends. Such DSLs are, however, too rigid, and do not adapt to changes in backends or to input graph properties or to both. We narrate our experiences in making an existing DSL, named Falcon, adaptive. The biggest challenge in the process is to not change the DSL code for specifying the algorithm. We illustrate the effectiveness of our proposal by auto-generating codes for vertex-based versus edge-based graph processing, synchronous versus asynchronous execution, and CPU versus GPU backends from the same specification.
\end{abstract}

%% file: introduction.tex
\section{Introduction}\label{sec:introduction}
Graphs model several real-world phenomena such as friendship, molecular interaction and co-authorship.
Several graph algorithms have been designed across domains to compute such relationships between entities.
Performance of these graph algorithms has become critical today due to the explosive growth of unstructured data.
For instance, to simulate a simple physical phenomenon, an algorithm may have to work with billions of particles.

On the other side, computer hardware is witnessing rapid changes with new architectural innovations.
Exploiting these architectures demands complex coding and good compiler support.
The demand intensifies in the presence of parallelization.
It is not uncommon to see a twenty-line textbook graph algorithm implemented using several hundred lines of optimized parallel code.

It would be ideal if a graph algorithm can be programmed at a high-level without worrying about the nuances of the hardware.
Domain-specific languages (DSLs) for graph  data analytics allow programmers to write complex algorithmic codes with minimal hardware knowledge and less programming effort.  The code generator of the DSL compiler  emits efficient parallel code~\cite{greenmarl,lighthouse,falcon}.
Such DSLs often disallow writing arbitrary programs, trading off generality for performance. This makes programming parallel hardware easy, and adapting to changes manageable.

In this work, we focus on a recent graph DSL named Falcon~\cite{falcon,dhfalcon}, which supports a wide variety of backends: CPU, GPU, multi-GPU, and distributed systems with CPU and GPU.  
It extends C language to allow graph processing being specified at a high-level.
Falcon provides special data types (such as \texttt{worklist} and \texttt{set}) as well as constructs (such as \texttt{foreach} and \texttt{reduction}) for simplifying algorithm specification and aiding efficient code generation.
Table~\ref{background:table1} compares various graph processing frameworks.

\begin{table}[b]
\centering
\scalebox{0.7}{
 \begin{tabular}{|r|c|crc|ccc|}
\hline
\small
	&	& \multicolumn{3}{c|}{Hardware Support}	& \multicolumn{3}{c|}{Iterators}\\
\multicolumn{1}{|c|}{Tool}	& DSL	& CPU	& GPU	& multi-GPU	& Edge	& Vertex	& Worklist\\
\hline
\hline
GreenMarl~\cite{Hong:2012:GDE:2150976.2151013} & $\surd$ & $\surd$ & $\times$& $\times$ & $\surd$& $\surd$& $\surd$\\
Galois~\cite{Pingali:2011:TPA:1993316.1993501} & $\times$ & $\surd$& $\times$& $\times$ & $\times$& $\times$& $\surd$\\
\textbf{Falcon}~\cite{falcon} & $\surd$ & $\surd$& $\surd$& $\surd$ & $\surd$& $\surd$& $\surd$\\
Totem~\cite{Gharaibeh:2012:YOT:2370816.2370866} & $\times$ & $\surd$& $\surd$& $\surd$ & $\times$& $\times$& $\times$\\
Gunrock~\cite{Wang:2016:GHG:3016078.2851145} & $\times$ & $\times$& $\surd$& $\times$ & $\times$& $\times$& $\surd$\\
LonestarGPU~\cite{nasre13:MAG:2517327.2442531} & $\times$ & $\times$& $\surd$& $\times$ & $\surd$& $\surd$& $\surd$\\
\hline
\end{tabular}
}
\caption{Comparison of different graph frameworks}
\label{background:table1}
\end{table}

Graph algorithms are challenging to parallelize due to their inherent \textit{irregularity}, which makes their data-access, control-flow, and communication patterns data-dependent.
For instance, vertex-based processing works well for road networks, but social networks demand an edge-based processing. Social networks have a skewed degree distribution and road networks have a large diameter.
Sequential processing of parallel loops demands synchronous execution, but independent loops can be more efficient with asynchronous processing.
Dense subgraphs can be efficiently processed using a topology-driven approach~\cite{nasre13}, whereas a data-driven worklist-based approach is better suited for sparse subgraphs.
Similarly, backend optimizations are quite different for different targets such as CPU and GPU.

Falcon, and other graph DSLs, allow writing code for a particular kind of processing. 
The code written in Falcon DSL needs modifications for an alternative way of processing.
Various syntactic elements in the program need to be changed for the alternative way. 
Thus, the code needs to be written separately for vertex-based and edge-based processing, for instance.
It would be ideal if one could generate different kind of code from the \textit{same DSL specification}.
Such a setup greatly simplifies the algorithmic specification, and also allows generating code for various situations / backends / graph types from the same specification.

\REM{
The burden, in this setup, shifts from DSL programmer to the compiler.
The only input that the programmer needs to specify (apart from the DSL code) is the type of processing required.
This can be easily specified via a command-line switch (e.g., \texttt{-vertex-based}, \texttt{-synchronous}, etc.), but without needing any modifications to the code.
This allows our compiler to generate vertex-based and edge-based graph processing code from the same algorithmic specification.
Similarly, it allows generating synchronous and asynchronous processing codes, wherein various iterations of the parallel processing are separated and not separated by a barrier respectively. 
The compiler is also equipped to generate CPU or GPU or multi-GPU code.
}

In this paper, we highlight the challenges faced in building such a versatile compiler.
In particular, we make the following contributions:
\begin{itemize}
\item We present a compiler which generates different implementations for the \textit{same DSL program} for a graph algorithm. In particular, the compiler can generate vertex-based or edge-based processing, synchronous or asynchronous codes, topology-driven versus data-driven processing, and CPU or GPU or multi-GPU codes.
\item We illustrate the effectiveness of the proposed compiler using several graph algorithms and several graphs of various types. The performance of the code generated with the proposed compiler is compared against other hand-tuned as well as generated codes.
\end{itemize}

\REM {
The rest of the paper is organized as below.
Chapter~\ref{sec:background} provides a brief background of Falcon DSL, as our work is implemented on top of it.
Chapter~\ref{sec:approach} presents the challenges faced in generating various codes from the same DSL.
Chapter~\ref{sec:results} evaluates the effectiveness of our approach using multiple frameworks, algorithms and graphs.
Chapter~\ref{sec:related} compares and contrasts the relevant related work,
and Chapter~\ref{sec:conclusion} concludes.
}

%% file: background.tex
\section{Falcon}\label{sec:background}

Falcon~\cite{falcon} is a Graph DSL which supports \CPU, \GPU and multi-GPU machines. It supports various data types, parallelization and synchronization constructs, and reduction operations. This makes programming graph analytic algorithms easy for heterogeneous targets. Falcon also supports dynamic graph algorithms. 
\par Falcon Graph DSL has data types {\tt Graph}, {\tt Point}, {\tt Edge}, {\tt Set} and {\tt Collection}. \texttt{Graph} stores a graph object, which consist of points and edges. Each {\tt Point} is stored as a {\tt union} of   {\tt int} and {\tt float}.
{\tt Edge} consists of source and destination points, and   {\it weight}. {\tt Set} is a static collection and implemented as a {\it Union-Find} data structure. The {\tt Collection} data type is dynamic and its size can vary at runtime. Elements can be added to  and deleted from a collection object at runtime.

 \par The {\tt foreach} statement is the parallelization construct of Falcon.
 It can be used to  iterate in different ways on different elements of graph object as  shown in Table~\ref{background:tabfalcon2}. 
{\tt Parallel} {\tt  Sections} statement of Falcon is used to write programs which use multiple devices of a machine at the same time. Falcon also supports reduction operations such as add ({\it RADD}) and mul ({\it RMUL}). It  has atomic library functions {\it MIN}, {\it MAX} etc., which are necessary for graph algorithms as they are {\it irregular}. 
The synchronization primitive of Falcon DSL is {\tt single} statement.
 It is a non-blocking lock and can be used to lock one element or a collection of elements, as shown in Table~\ref{background:tabfalcon1}.

\begin{table}
 \tiny{
\centering
\begin{tabular}{ |p{1cm} |p{1cm} |p{5cm}| }
 \hline
  Data Type &Iterator  & Description  \\
\hline
  Graph   &  points  &  Iterate over  all points in graph \\
Graph    & edges    & Iterate over all edges in graph \\
Point   & nbrs     & Iterate over all neighboring points (Undirected {\tt Graph})\\
Point   & innbrs     & Iterate over all src point of incoming edges\\
Point   & outnbrs   & Iterate over dst   point of   outgoing  edges \\
Set     &      & Iterate over all items in a Set \\
Collection       &    & Iterate over all items in a Collection\\
\hline
\end{tabular}
\caption{ {\tt foreach}  statement iterators in Falcon}
\label{background:tabfalcon2}
}
\tiny{
\centering

 \begin{tabular}{|l|l|}
\hline
\shortstack{ \textbf{single}(t1) \{stmt block1\}\\ \textbf{else} \{stmt block2\}}  &\shortstack{ The thread that gets a lock on  item t1\\ executes  stmt block1 and other threads\\ execute stmt block2.}\\
\hline
\shortstack{ \textbf{single}(coll) \{stmt block1\}\\ \textbf{else} \{stmt block2\}}  & \shortstack{The thread that gets a lock on  all elements\\ in the collection executes  stmt block1\\ and others execute stmt block2.}\\
\hline
 \end{tabular}
\caption{{\tt single} statement (synchronization) in Falcon}
\label{background:tabfalcon1}
}

\end{table}
A graph object can be processed in multiple ways in Falcon. This leads to the flexibility of the same algorithm being specified in different ways.   A programmer can iterate over {\it edges} of a graph object and then extract the source ({\it src}) and the destination ({\it dst}) points of each edge. Another method is to iterate over all {\it points} of the graph object. 
Then for each point, the processing can iterate over {\it outnbrs} or {\it innbrs}.
This is illustrated in Algorithms~\ref{background:algo1} and \ref{background:algo2}.
\begin{algorithm}[t]
\scriptsize
\SetKwProg{Fn}{}{ \{}{\}}
\fontsize{8.2pt}{5pt}\selectfont{
\SetAlgoLined
int  changed = 0;  // Global variable \label{line:globdecl}}\\
\Fn(){\textbf{relaxgraph}(Point  p, Graph  graph)} {
                        \textbf{foreach} (t In p.outnbrs)\\
        \hspace{0.05in} MIN(t.dist, p.dist + graph.getweight(p, t), changed);    \label{line:modidist}\\
}
\Fn(){\textbf{main}(int argc, char *argv[])} {
        Graph graph;    \\
        graph.addPointProperty(dist, int);      \label{line:add-dist}\\
        graph.read(argv[1]);            \label{line:readgraph}  \\
        //make {\it dist} infinity for all points.\\
        \textbf{foreach} (t In graph.points)t.dist=MAX\_INT;     \label{line:infinity}\\
        graph.points[0].dist = 0;       // source has dist 0    \label{line:initsource} \\
        \While {(1)}{     
         changed = 0;           \\\label{line:initchanged}
                \textbf{foreach} (t In graph.points)  relaxgraph(t,graph);\label{line:relaxfun}\\
                if (changed == 0) break;        //reached fix point\label{line:checkexit}\\

        }\label{line:ssspendlopp}

        }
\caption{SSSP: iterating over Points in Falcon}
\label{background:algo1}
        \end{algorithm}
\begin{algorithm}[H]
\scriptsize
\SetKwProg{Fn}{}{ \{}{\}}
\fontsize{8.2pt}{5pt}\selectfont{
\SetAlgoLined
int  changed = 0;  // Global variable \label{line:eglobdecl}}\\
\Fn(){\textbf{relaxgraph}(Edges  e, Graph  graph)} {
        Point (graph)p,(graph)t;\\
    p=e.src;\\
    t=e.dst;\\
    MIN(t.dist, p.dist + e.weight, changed);    \label{line:emodidist}\\
}
\Fn(){\textbf{main}(int argc, char *argv[])} {
        Graph graph;    \\
        graph.addPointProperty(dist, int);      \label{line:eadd-dist}\\
        graph.read(argv[1]);            \label{line:ereadgraph}  \\
        //make {\it dist} infinity for all points.\\
        \textbf{foreach} (t In graph.points)t.dist=MAX\_INT;     \label{line:einfinity}\\
        graph.points[0].dist = 0;       // source has dist 0    \label{line:einitsource} \\
        \While {(1)}{     
         changed = 0;           \\\label{line:einitchanged}
                \textbf{foreach} (e In graph.edges)  relaxgraph(e,graph);\label{line:erelaxfun}\\
                if (changed == 0) break;        //reached fix point\label{line:echeckexit}\\

        }\label{ssspendlopp}

        }
\caption{SSSP: iterating over Edges in Falcon}
\label{background:algo2}
        \end{algorithm}

Both the algorithms are for  Single Source Shortest Path (SSSP) computation. It computes the shortest path from source point (point zero) to all other points in the graph object. 
 In Algorithm~\ref{background:algo1}, the processing is done using {\it points} (Line~\ref{line:relaxfun}) and {\it outnbrs} (Line~\ref{line:modidist}) iterators. In Algorithm~\ref{background:algo2}, the computation is performed using {\it edges} (Line~\ref{line:erelaxfun}) iterator.
 In both the algorithms all the edges $t\rightarrow p$ in the graph object are considered. Then {\it dist} value of point {\it t} is reduced to {\it Min(t.dist, p.dist+ weight($p\rightarrow t$))} using the atomic function {\it MIN}.
 If there is any change in the value of {\it t.dist}, the variable {\it changed} is set to one. The computation stops when the value of {\it dist} does not change for any point in the graph object. Performance of an algorithm depends on the graph structure, hardware architecture, etc.
 Algorithm~\ref{background:algo1} may perform well over Algorithm~\ref{background:algo2} for one input graph, but may not for another, \REM{ input graph} on the same hardware architecture. This depends upon several  properties such as the degree of vertices and diameter of the graph object etc. Similary the worklist (\texttt{Collection}) based code also needs to be coded separately in Falcon.

Such a flexible processing is an artifact of \textit{irregular} algorithms (such as graph algorithms) wherein the data-access pattern, the control-flow pattern, as well as the communication pattern is unknown at compile time, as all are dependent on the graph input.
 Thus, it is difficult to identify which method would be suitable for an algorithm:  it depends on the graph object.

The random graphs (Erd$\ddot{o}$s R$\acute{e}nyi$ model) typically perform well with iterating over {\it points}. The social and rmat graphs which follow power-law degree distribution~\cite{Gharaibeh:2012:YOT:2370816.2370866} are benefited mostly by iterating over {\it edges}, especially on \GPU devices.  Power-law degree distribution indicates huge variance in degree distribution of the vertices. This can result in thread-divergence in \GPU, when parallelized over {\it points} and iterated over their {\it outnbrs} or {\it innbrs}.  Road networks benefits with worklist based processing on \CPU.

Our goal in this work is to bridge the gap between easy DSL specification and versatility in generating various kinds of codes.
Thus, from the same Falcon specification, we want to generate vertex-based or edge-based OpenMP or CUDA codes.

%% file: approach.tex
\begin{algorithm}[t]
\small
\SetKwProg{Fn}{}{ \{}{\}}
int   changed = 0, lev = 0;\label{appr:algo:bfs:decl}\\
void \Fn{ \textbf{BFS}(Point p, Graph graph, int lev)}{
	\ForEach{t In p.outnbrs}{\label{appr:algo:bfs:outnbr}
		\IfBlock{t.dist $>$ lev + 1}{\label{appr:algo:bfs:update}
			t.dist = lev + 1;\\
			changed = 1;\\
		}\label{appr:algo:bfs:updateend}
	}
} 
int \Fn{ \textbf{main}(int argc, char *argv[])}{
  Graph graph;\\
	graph.addPointProperty(dist, int);\label{appr:algo:bfs:distppty}\\
	graph.read(argv[3]);\label{appr:algo:bfs:read}\\
foreach(t In graph.points) t.dist=1234567890;\label{appr:algo:bfs:reset}\\
	graph.points[0].dist = 0;\label{appr:algo:bfs:initsrc}\\
	\While{1}{\label{appr:algo:bfs:whilebegin}
		changed = 0;\label{appr:algo:bfs:initch}\\
foreach(t In graph.points)(t.dist == lev) BFS(t, graph, lev);\label{appr:algo:bfs:call}\\
		if (changed == 0) break;\label{appr:algo:bfs:checkexit}\\
		lev++;\label{appr:algo:bfs:incrlev}\\
	}\label{appr:algo:bfs:whileend}
}
\caption{BFS Algorithm in Falcon for \CPU}
\label{appr:algo:bfs}
\end{algorithm}


\section{Our Approach}\label{sec:approach}
Algorithm~\ref{appr:algo:bfs} shows the Falcon DSL code for Breadth First Search (BFS)  computation.
\REM{
The graph object is added with a property {\it dist} (Line~\ref{appr:algo:bfs:distppty} ), which is used to store the bfs distance.
The graph object is read and then bfs distance of all vertices  is made infinity (Lines~\ref{appr:algo:bfs:read} - \ref{appr:algo:bfs:reset} ).
 Then the bfs distance of the source vertex is made zero (Line ~\ref{appr:algo:bfs:initsrc} ). The bfs distance is computed in the {\tt while} loop (Lines~\ref{appr:algo:bfs:whilebegin} - \ref{appr:algo:bfs:whileend} ).
 A parallel {\tt foreach} call to the {\it BFS()} function is made ( Line~\ref{appr:algo:bfs:call} ). The {\it BFS()} algorithm uses an atomic free, level based traversal.
 The {\it lev} variable is initialized to zero (Line~\ref{appr:algo:bfs:decl} ).
 The  {\it lev} variable is incremented by a value of one (Line~\ref{appr:algo:bfs:incrlev} ) at the end of each iteration of the {\tt while} loop. 
Inside the {\it BFS()} function, for each edge $p\rightarrow t$, if {\it t.dist} is greather than {\it lev+1}, then {\it t.dist} is updated and {\it changed} variable is made one (Lines~\ref{appr:algo:bfs:update} - \ref{appr:algo:bfs:updateend} ).
 The {\it changed} variable is set to zero (Line ~\ref{appr:algo:bfs:initch} ) before calling {\it BFS()}. 
The algorithm terminates when there is no update of {\it dist} value for any of the vertex during the invocation of {\it BFS()} function.
 When such a condition happens, the value of the changed variable will remain zero and the exit condition is satisfied (Line ~\ref{appr:algo:bfs:checkexit} ).  
}
The algorithm is vertex-based and uses {\it points} (graph nodes) and {\it outnbrs} (of each node) iterators.
The above Falcon program is for \CPU. 
A programmer has to write separate programs for \GPU which will have $\textless GPU\textgreater$ in the DSL code~\cite{falcon}. 
Similarly, different programs need to be written for edge-based or worklist-based codes.
It would be ideal if the programmer simply specifies \textit{what} rather than \textit{how}, and the DSL compiler takes care of the appropriate code generation.
We support it in this work.
In our proposal, the programmer needs to specify simply different compile-time arguments. 
This triggers generation of parallel C++ (CUDA) code  matching the output of the {\it edge}, {\it vertex} or {\it worklist} based BFS, targeting \CPU  (\GPU respectively), from  a single DSL code. 
We explain each of these transformations in the subsections below.

\subsection{Vertex-based versus Edge-based}\label{sec:vertexedge}
Edge-based processing improves load-balance, while vertex-based codes improve propagation of information across the graph and can also reduce synchronization requirements.
Conversion of vertex-based to edge-based and vice-versa are done completely at the abstract-syntax tree (AST) level  by traversing the AST and modifying its eligible parts. An important conversion non-triviality stems from the edge-based processing being a single loop, while the corresponding vertex-based processing is a nested loop (outer loop over vertices, and inner loop over all the neighbors of each vertex).
Pseudo-code for the vertex-based to edge-based transformation is presented in Algorithm~\ref{codege:optcomm}. The other way is similar.

\begin{algorithm}[t]
\scriptsize
\KwIn{Falcon vertex-based DSL code}
\KwOut{C++/CUDA edge-based code based on the command-line}
 \textbf{ begin  Step1:-} Mark  outermost {\tt foreach} statement (Done by  Falcon Parser).\\
 \lIf {  statement.type == {\tt foreach} \&\& level == 0} {
   t.outer = {\tt true}
}
 \lIf {  statement.type == {\tt foreach} \&\& level == 1} {
   t.outer = {\tt false}
}
 \textbf{end Step1}\\
 \textbf{begin Step2:-}  Convert vertex code to edge code \\
  \ForAll { {\tt foreach} statement $f1$ in $program$ } {
 \If { $f1$.outer == {\tt true} \&\& $f1$.iterator == $points$}{
  \ForAll { {\tt foreach} statement $f2$ in $program$ } {
  \If{$f2.def.fun == f1.call.fun$ \&\& $f2.itr == outbrs || f2.itr == innbrs$}{
  modify iterator of $f1$  to {\tt edges}\\
   modify  $1^{st}$ argument to {\tt Edge} in    $f2.def.fun$\\
  create  $f2.itr$ and $f1.itr$ in  $f2.def.fun$ using {\tt Edge}\\
   remove {\tt foreach} in $f2.def.fun$\\
 // generate code (Done by Falcon)
  }
} } } 
\textbf{end Step2}\\
\REM{
  \textbf{begin Step3:-} Convert edge code to vertex code\\
  \ForAll { {\tt foreach} statement $f1$ in $program$ } {
 \If { $f1$.outer=={\tt true} \&\& $f1$.iterator==$edges$}{
  \textbf{function} $fun$ =$f1.call.fun$ \\
  modify iterator of $fun$  to points\\
   modify $1^{st}$ argument of fun to {\tt Point} ({\it p})\\
   insert $outnbrs$ or $innbrs$  iterator in fun using {\tt Point} ({\it t})\newline
\shortstack{   replace edge in $f1.call.fun$ by \\ \hspace{0.3in}$p\rightarrow t$ (outnbrs) or $t\rightarrow p$ (innbrs)}\\
 // generate code (Done by Falcon)\\
  }
}
\textbf{end Step3}
}
  \caption{Code transformation for vertex based code}
 \label{codege:optcomm}
 \end{algorithm}
In vertex-based to edge-based conversion,  the subtree is eligible for conversion if:
(i) the subtree is rooted at a function node whose only child is a node for a \texttt{foreach} which iterates through a point's neighbors, and
(ii) the function is the only statement in the body of a \texttt{foreach} which iterates through the graph-points. 
Once the eligible parts are found, we switch the points iterator of the \texttt{foreach} statement from which the function is called to edge iterator, and then remove the \texttt{foreach} statement in the function. 
The conversion also requires change in the function's signature as its argument was earlier a point, while now it is an edge. 
It also necessitates defining two new variables at the beginning of the function corresponding to the source and the destination of the edge. 
The name of one of the two variables is the name of the point which was the former parameter of the function. 
The other variable's name is derived from the iterator of the \texttt{foreach} statement removed from the function earlier. 
The order in which these names are mapped to the variables depends on the iterator used in the removed \texttt{foreach}. 
If the iterator is over out-neighbors, the name of the iterator is mapped to the destination vertex of the edge.
Otherwise, we map it to the source vertex. 
Such an implementation allows the rest of the processing in the iteration to be arbitrary, and reduces the number of alterations the compiler needs to perform to the underlying code.

\REM{
In edge-based to vertex-based conversion, the compiler needs to do the opposite (see Algorithm~\ref{codege:optcomm}). 
Here, it finds the \texttt{foreach} statement iterating through the edges of a graph which contains a function call statement as the only statement in the loop-body. 
For such a \texttt{foreach}, the iterator is changed to iterate over points and a new \texttt{foreach} statement iterating through the neighbors of the point is introduced in the called function (which encloses the body of the called function). 
This conversion necessitates the parameter of the called function to be changed from edge type to point type. 
It also requires defining a new variable which represents the edge between the point passed as a parameter and the iterator which represents the point's neighbor. 
Such an implementation also allows the rest of the processing in the iteration to be arbitrary.
}

An important artifact of this processing conversion is that it affects the way graph is stored in memory.
In vertex-based code, Falcon (and other frameworks) store the graph in compressed sparse-row (CSR) format. 
Compared to edge-list representation, CSR format reduces the storage requirement and allows quick access to a vertex's out-neighbors.
In edge-based codes, on the other hand, the graph is stored in edge-list format (\textsf{source destination weight}) which enables quick retrieval of the source and the destination points of an edge.
\REM{
However, one graph format also forbids the other format's advantages.
For instance, if the graph is stored in CSR format, retrieving an arbitrary edge (say, edge $p \rightarrow q$) is time-consuming (requires a search over the out-neighbors of $p$).
On the other hand, if the graph is stored as edge-list, retrieving an edge, given the source and the destination vertices is even more time-consuming, as the edges are not indexed on vertices. 
In many cases (and all in our testbed), accessing edges is not arbitrary. 
Edges are accessed in a particular order, for instance, iterating through out-neighbors of a point. 
So edge between a point and its out-neighbor can be retrieved quickly if the graph is stored in CSR format. 
Unfortunately, while changing vertex-based code to edge-based code, accessing edges in this manner becomes time-consuming. 
To overcome this, the code where the edge is retrieved is replaced by the edge passed as an argument to the transformed function. 
This resolves the issue of enumerating over neighbors.
}

\subsection{Synchronous versus Asynchronous}\label{sec:syncasync}
By default, Falcon generates BSP-style synchronous code, that is, it inserts a barrier at the end of a parallel construct. 
While this works well in several codes and eases arguing about the correctness (due to data-races restricted to within-iteration processing across threads), synchronous processing may be overly prohibitive in certain contexts.
Especially, in cases where processing across iterations is independent and the hardware does not necessarily demand single-instruction multiple data (SIMD) execution, asynchronous processing may improve performance.
Arguing about the correctness-guarantees gets so involved with asynchrony, that some DSLs enforce synchronous-only code generation.

\begin{algorithm}[t]
  \scriptsize
  \KwIn{CFG of the function where kernels are launched.}
  \KwOut{None}
  
  \ForEach{Node nd \textbf{in} cfg} {
    nd.visited = 0\\
    nd.barrier = False\\
    nd.predecessor\_count = 0
  }
  BFS\_Mod(cfg.root)// set predecesor\_count of nodes by running bfs from root of the CFG\\
  parallelize(cfg.root, None) // Algorithm~\ref{codegen:async}
  \caption{Driver code to generate asynchronous code}
  \label{codegen:driver}
\end{algorithm}

\begin{algorithm}[t]
  \scriptsize
  \SetKwProg{Fn}{}{ \{}{\}}
  
  \Fn(){\textbf{parallelize}(Node node, Node knode)} {
    \eIf{node.stmt.type == KERNEL\_LAUNCH}{
      rset = get\_read\_set\_from\_function(node.stmt.function)\\
      wset = get\_write\_set\_from\_function(node.stmt.function)\\
      \If{knode != None} {
        \eIf{rset $\cap$ knode.wset != EMPTY $\parallel$ wset $\cap$ knode.rset != EMPTY $\parallel$ wset $\cap$ knode.wset != EMPTY} {
          knode.barrier = True
        } {
          node.rset = node.rset $\cup$ knode.rset\\
          node.wset = node.wset $\cup$ knode.wset
        }
      }
      knode = node
    } {
      \If{knode != None} { 
        rset = get\_read\_set\_from\_statement(node.stmt)\\
        wset = get\_write\_set\_from\_statement(node.stmt)\\
        \If{rset $\cap$ knode.wset != EMPTY $\parallel$ wset $\cap$ knode.rset != EMPTY $\parallel$ wset $\cap$ knode.wset != EMPTY} {
          knode.barrier = True\\
          knode = None
        }
      }
    }
    node.visited += 1\\
    \If{node.predecessor\_count == 0 $\parallel$ node.visited == node.predecesor\_count}{
        \lForEach{Node nd \textbf{in} node.successors}{
          parallelize(nd, knode)
        }
    }
  }
  \caption{Mark node as barrier/barrier-free}
  \label{codegen:async}
\end{algorithm}

Our proposal is to allow the programmer to generate synchronous or asynchronous code without having to change the algorithm specification code in the DSL.
Achieving this necessitates identifying independent processing in the code. 
Towards this, we maintain read and write sets of global variables and the graph attributes used in each \textit{target function} separately. A \textit{target function} is a function which is being called in the body of a \texttt{foreach} statement and the function call is the only statement inside the body of the \texttt{foreach}.  On CPU, it is the parallel loop body, while on GPU, this function becomes the kernel.

The code conversion has two steps, as shown in Algorithms~\ref{codegen:driver} and \ref{codegen:async}. In Step 1, we mark nodes in the control-flow graph (CFG); and in Step 2, we generate the appropriate code.
In Step 1, we construct the CFG of the \textit{target function} call.
Using the read and the write sets corresponding to each of the target functions, we mark each node of the CFG as \textit{barrier-free} or not. A barrier-free node signifies that the target function corresponding to the barrier-free node can be executed concurrently with the children of this node.
A node is barrier-free if: 
(i) there is no dependency between the node and each of its children in CFG. 
(ii) there is no dependency between the node and the codes between the node and its children.

If a node is {\it barrier-free}, we pass the read and the write sets of the node to its children. 
We do this so that the grand-child should not have any dependency with the grand-parent node to declare its parent {\it barrier-free} (and so on).
We follow this process to mark all the CFG nodes in breadth-first search order.

In Step 2, based on the target code the programmer wants, different procedures are followed to make the code asynchronous.
If the target is GPU, all the nodes marked as {\it barrier-free} do not contain a barrier \textit{cudaDeviceSynchronize()} after the kernel launch. 
Also, each of the {\it barrier-free} kernels is launched in different streams of the same GPU.
On the other hand, if the target is CPU, the \textit{target function} call corresponding to the {\it barrier-free} node is put in a section of an OpenMP parallel region, and its children and the code between the node and its children in another section.
The compiler then recursively checks if the child nodes are barrier-free or not. 
If they are, then a new OpenMP \texttt{parallel sections} construct is introduced inside the section where the child was put in earlier, because of its {\it barrier-free} parent node.
This recursive introduction of parallel sections continues until a {\it non-barrier-free} node is found, or until the processing reaches the end of the function where these \textit{target functions} are called. 
The introduction of OpenMP constructs is done by adding new nodes in the AST.
For a parallel region, two nodes are added: one each for the start and the end of the construct. In a similar manner, for each section, a node for the start and another node for the end is added. 
Adding these nodes is easy if both the node and its children in the CFG lie in the same scope. 
We can then simply add a node prior to and another node right after the barrier-free node. 
The processing gets involved when a node and its children are in different scopes.
In such cases, we need to find the predecessors of these nodes which lie in the same scope.

\REM{
\textbf{Discussion:} The way we group two parallel constructs may not lead to optimal grouping. 
Finding the optimal grouping is not feasible as it depends on the time required to execute the parallel constructs. 
For instance, $A$, $B$, and $C$ are parallel constructs where $A$ and $B$ are independent, $B$ and $C$ are independent, and $A$ and $C$ are dependent. 
If the time required to execute $A$ is less compared to $B$ and $C$, then grouping $A$ and $B$ does not lead to optimal performance.
On the other hand, if the time required to execute $C$ is less compared to $A$ and $B$, then grouping $A$ and $B$ does lead to optimal performance.
}

\subsection{Data-driven versus Topology-driven Processing}\label{sec:datatopo}
It is a two step process. In step one, each kernel is checked if it can be converted into worklist-based and in step two, AST is modified such that code generation module generates worklist based code.
The primary program analysis required in this transformation is to identify where point attributes are getting modified (e.g., distance of nodes in SSSP) and push such points (vertices) into the worklist.
Our method goes through the AST of \texttt{foreach} and checks for such attribute updates and checks.
Based on the target architecture, code generation module generates worklist based code for the particular target. In case of CPU, Falcon generates Galois worklist based code. In case of GPU, Falcon library provides a worklist interface.

\subsection{CPU, GPU and Multi-GPU Codes}\label{sec:cpugpu}
Falcon~\cite{falcon} requires a programmer to write different DSL code for different backends.
It uses $<$GPU$>$ tag to specify a GPU variable.
Falcon compiler generates GPU code if there exists a GPU variable in the program and converts function to GPU kernel if one of the parameters is a GPU variable.
We modified the Falcon grammar so that compiler does not need a GPU tag.
It recognizes a device-independent version of the DSL code.
Based on the command-line argument, our compiler generates code for an appropriate target device.

The compiler generates the GPU code in the following manner.
First, it marks all the \textit{target functions} as kernels.
Second, it marks the global variables used in the \textit{target functions} as GPU variables.
Third, it makes a GPU copy of each of the variables of type graph, set and collection.
Fourth, it replaces the CPU copy of a graph, set or collection with its corresponding GPU copy.

To generate multi-GPU code, the programmer has to use \texttt{parallel} {\tt sections} construct of Falcon.
The Falcon compiler assumes that each of the sections is independent of each other.
We identify the number of sections in a parallel sections construct and map each of the sections to a different GPU.
For each graph, set and collection used in a particular section, a GPU copy is created in the mapped GPU.
It may happen that the programmer has used a single graph and used that graph in multiple sections.
In such cases, the graph needs to be copied to each GPU.
For each of those GPU copies, we keep track of the attributes of the graph or its points/edges used in the target functions where the graph is passed as an argument.
This helps us to replace the graph whose attribute is accessed in CPU by the appropriate GPU copies where the accessing attribute is present.
Now if an attribute of a GPU graph is accessed in the CPU, the Falcon compiler generates a call to \texttt{cudaMemcpy} to copy the attribute from GPU to CPU or from CPU to GPU based on whether the programmer has read or written to the attribute.
One advantage of assuming independent sections is that the attributes accessed in CPU can be changed on maximum one GPU,
which eases our analysis and code generation.

\REM{
\subsection{Data-Transfer Aggregation Optimization}\label{sec:optimizations}
In Falcon, the custom properties of points/edges in a graph object is stored as an array where $n^{th}$ position stores the value corresponding to point/edge $n$.
If a property of a point/edge of a GPU graph object is read in the CPU, a data transfer from GPU to CPU is necessary.
By default, for $n$ such reads, $n$ instances of data transfer are required.
This is not always desirable and can quickly become a performance bottleneck due to slow PCI-e across the devices. 
To overcome this, we combine multiple such data transfers into a single data transfer, where feasible.

Such an optimization is particularly beneficial for \texttt{for} and \texttt{while} loops.
If the instructions inside the body of a \texttt{for} or {\tt while} loop only read a particular property (and may write to other properties), we copy the array related to that property to CPU before the beginning of the loop using a single data transfer instruction. 
The instructions in the loop then use the CPU copy.
This often improves the performance as a loop is expected to be executed several times.
}

%% file: results.tex
\section{Experimental Evaluation}\label{sec:results}
We modified Falcon's~\cite{falcon} abstract syntax tree (AST) processing and code generation to generate various types of codes presented in the last section.
For CPUs, it generates parallel code with OpenMP for vertex and edge based processing, while Galois worklist code for worklist based processing. For GPU and multi-GPU targets, it generates CUDA code.
\subsection{Experimental Setup}\label{expt:setup}
We used a range of graph types to assess the effectiveness of our proposal.
The dataset graphs in our experimental setup and their characteristics are presented in Table~\ref{expt:chars}.
We used four graph algorithms in our testbed: Breadth-First Search (BFS), Connected Components (CC), Minimum Spanning Tree computation (MST) and Single-Source Shortest Paths computation (SSSP).
All these algorithms are fundamental in graph theory and form building blocks in various application domains.
We compare the generated codes against the following frameworks: Galois~\cite{Pingali:2011:TPA:1993316.1993501}, Totem~\cite{Gharaibeh:2012:YOT:2370816.2370866}, Green-Marl~\cite{Hong:2012:GDE:2150976.2151013}, LonestarGPU~\cite{nasre13:MAG:2517327.2442531} and Gunrock~\cite{Wang:2016:GHG:3016078.2851145}.
Our auto-generated codes perform similar to hand-optimized Falcon codes.
Therefore, in the sequel, we discuss directly our proposed techniques embedded into existing Falcon, unless otherwise stated.

The CPU benchmarks for OpenMP are run on an Intel XeonE5-2650 v2 machine with 32 cores clocked at 2.6 GHz with 100GB RAM, 32KB of L1 data cache, 256KB of L2 cache and 20MB of L3 cache. The machine runs CentOS 6.5 and 2.6.32-431 kernel, with GCC version 4.4.7 and OpenMP version 4.0. The CUDA code is run on Tesla K40C devices each having 2880 cores clocked at 745 MHz with 12GB of global memory. Eight similar GPU devices are connected to the same CPU device.

\begin{table}
\parbox{0.4\textwidth}{
\centering
\footnotesize
\begin{tabular}{|l|r|r|r|}
\hline
Graph	& \#nodes	& \#edges	& max-\\
	& $\times10^6$	& $\times10^6$	& degree\\
\hline
USA-CTR	& 14	& 34 	& 9    \\
USA-full	& 24	& 58 	& 9    \\
orkut	& 3	& 234 	& 33313 \\
sinaweibo	& 21	& 261 	& 278491 \\
rand-25M	& 25	& 100 	& 17   \\
rand-50M	& 50	& 200 	& 18   \\
rand-75M	& 75	& 300 	& 18   \\
rand-100M	& 100	& 400 	& 18   \\
rand-125M	& 125	& 500 	& 19   \\
rmat-10M	& 10	& 100 	& 1873 \\
rmat-20M	& 20	& 200 	& 2525 \\
rmat-40M	& 40	& 400 	& 3333 \\
rmat-50M	& 50	& 500 	& 4132 \\
\hline
\end{tabular}
\caption{Benchmark characteristics}
\label{expt:chars}
}
\hfill
\parbox{0.5\textwidth}{
\centering
\footnotesize
\begin{tabular}{|l|r|r|r|r|}
\hline
Graph	 	& BFS	& CC	& MST	& SSSP \\
\hline
USA-CTR	        & 3456	& 14103	& 727	& 30299 \\
USA-full	& 9113	& 24061	& 779	& 72857 \\
orkut	       & 269	& 623	& 5509	& 267   \\
sinaweibo	& 1234	& 1955	& 30556	& 1151  \\
rand-25M	& 131	& 411	& 2832	& 561   \\
rand-50M	& 270	& 823	& 6665	& 1142  \\
rand-75M	& 414	& 1416	& 11265	& 1796  \\
rand-100M	& 583	& 2192	& 15860	& 2413  \\
rand-125M	& 756	& 2909	& 25973	& 3194  \\
rmat-10M	& 133	& 387	& 2770	& 536   \\
rmat-20M	& 266	& 789	& 6240	& 1169  \\
rmat-40M	& 542	& 1601	& 13232	& 2405  \\
rmat-50M	& 707	& 2026	& 16825	& 3808  \\
\hline                                   
\textbf{Total}		& 17874	& 53296	& 139233&  121598 \\
\hline
\end{tabular}
\caption{Baseline times (ms) of Falcon on GPU}
\label{expt:baselines}
}
\end{table}
\subsection{Baselines and Comparison with Other Frameworks}\label{expt:comparison}
The baseline execution times of Falcon on GPU are listed in Table~\ref{expt:baselines}.
We observe that the execution times on road networks are particularly high for propagation based algorithms such as BFS, SSSP and CC.
This occurs because unlike other graphs, road networks have large diameters, leading to many iterations of the algorithm with parallelism not enough for GPU.
The opposite occurs for MST.

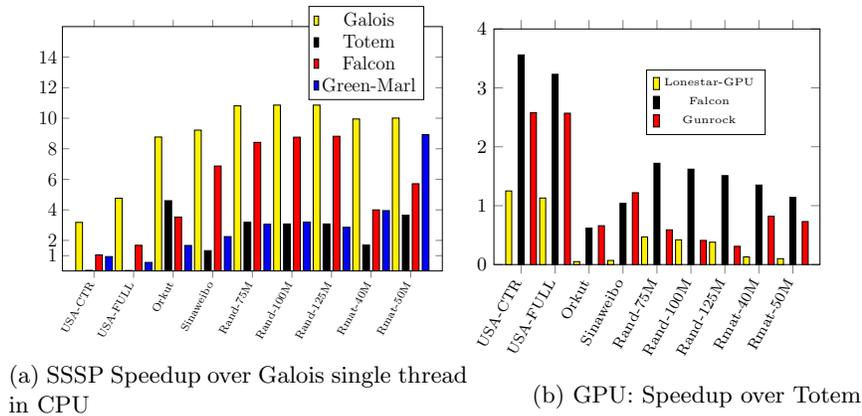
\begin{figure*}
\begin{subfigure}{0.5\linewidth}
\centering
  \begin{tikzpicture}[scale=0.6]
    \begin{axis}[
        ybar,
        ymin=0, ymax=16,
        height=7.0cm,
        width=10cm,
        ytick={ 1,2,4,6,8,10,12,14},
        x tick label style={
        rotate=60,anchor=east,font=\scriptsize},
        symbolic x coords={
          USA-CTR, USA-FULL, Orkut,Sinaweibo,Rand-75M,Rand-100M,Rand-125M,Rmat-40M,Rmat-50M
        },
        xtick=data,
        bar width=0.15cm,
        xticklabel style={rotate=1},
        legend style={
          at={(0.8,0.7)},
          anchor=south,
          legend columns=1,
          /tikz/every even column/.append style={font=\large}
        },
        legend entries={Lonestar-GPU,Falcon,Gunrock},
        legend image code/.code={%
      \draw[#1] (0cm,-0.1cm) rectangle (0.1cm,0.1cm);
    },  
    every node near coord/.append style={color=black, rotate=90, anchor=south, font=\large},
             every axis/.append style={font=\large},
    ]
    \addplot [ fill=yellow] coordinates {
      (USA-CTR,3.18)
      (USA-FULL,4.76) 
      (Orkut, 8.77)
      (Sinaweibo,9.22)
      (Rand-75M,10.81) 
      (Rand-100M,10.86) 
      (Rand-125M,10.86) 
     (Rmat-40M,9.95) 
      (Rmat-50M,10.01) 
            };
    \addplot [ fill=black] coordinates {
      (USA-CTR,0.04)
      (USA-FULL,0.03) 
      (Orkut, 4.60)
      (Sinaweibo,1.33)
      (Rand-75M,3.19) 
      (Rand-100M,3.08) 
      (Rand-125M,3.08) 
      (Rmat-40M,1.7) 
      (Rmat-50M,3.65) 
            };

    \addplot [ fill=red] coordinates {
      (USA-CTR,1.05)
      (USA-FULL,1.68) 
      (Orkut, 3.53)
      (Sinaweibo,6.87)
      (Rand-75M,8.42) 
      (Rand-100M,8.76) 
      (Rand-125M,8.82) 
      (Rmat-40M,4.00) 
      (Rmat-50M,5.71) 
            };
    \addplot [ fill=blue] coordinates {
      (USA-CTR,0.94)
      (USA-FULL,0.56) 
      (Orkut, 1.67)
      (Sinaweibo,2.25)
      (Rand-75M,3.06) 
      (Rand-100M,3.2) 
      (Rand-125M,2.86) 
      (Rmat-40M,3.95) 
      (Rmat-50M,8.93) 
    };
    \legend{\large{Galois},\large{Totem}, \large{Falcon},\large{Green-Marl}}
    \end{axis}
  \end{tikzpicture}
  \caption{SSSP Speedup over Galois single thread in CPU}
  \label{expt:ssspcpucompare}
\end{subfigure}%
  \begin{subfigure}{0.5\textwidth}
   \begin{tikzpicture}[scale=0.8]
    \begin{axis}[
         ybar=3pt,
        ymin=0, ymax=4,
        height=5.5cm,
        width=7cm,
x tick label style={
rotate=60,anchor=east,font=\scriptsize},
        symbolic x coords={
           USA-CTR, USA-FULL, Orkut,Sinaweibo,Rand-75M,Rand-100M,Rand-125M,Rmat-40M,Rmat-50M
            },
       xtick=data,
        bar width=0.1cm,
         xticklabel style={rotate=1},
       legend style={
            at={(0.65,0.55)},
            anchor=south,
            legend columns=1,
           /tikz/every even column/.append style={font=\tiny}
        },
        legend entries={Lonestar-GPU,Falcon,Gunrock},
        legend image code/.code={%
      \draw[#1] (0cm,-0.1cm) rectangle (0.1cm,0.1cm);
   },  
 every node near coord/.append style={color=black, rotate=90, anchor=south, font=\tiny},
             every axis/.append style={font=\small},
    ]
    \addplot [ fill=yellow] coordinates {
      (USA-CTR,1.25)
      (USA-FULL,1.13) 
      (Orkut, 0.05)
      (Sinaweibo,0.07)
      (Rand-75M,0.47) 
      (Rand-100M,0.42) 
      (Rand-125M,0.38) 
      (Rmat-40M,0.13) 
      (Rmat-50M,0.10) 
            };
    \addplot [ fill=black] coordinates {
      (USA-CTR,3.56)
      (USA-FULL,3.23) 
      (Orkut, 0.62)
      (Sinaweibo,1.04)
      (Rand-75M,1.72) 
      (Rand-100M,1.62) 
      (Rand-125M,1.51) 
      (Rmat-40M,1.35) 
      (Rmat-50M,1.14) 
            };
    \addplot [ fill=red] coordinates {
      (USA-CTR,2.58)
      (USA-FULL,2.57) 
      (Orkut, 0.66)
      (Sinaweibo,1.22)
      (Rand-75M,0.59) 
      (Rand-100M,0.41) 
      (Rand-125M,0.31) 
      (Rmat-40M,0.82) 
      (Rmat-50M,0.73) 
            };
 \legend{Lonestar-GPU,Falcon, Gunrock}
 \end{axis}
   \end{tikzpicture}
\caption{GPU: Speedup over Totem}
\label{expt:ssspgpucompare}
\end{subfigure}%
\caption{SSSP comparison}
\label{expt:ssspcpugpu}
\vspace{0.1in}
\end{figure*}

\begin{figure*}[h]
\centering
\begin{subfigure}{0.5\linewidth}
\centering
   \begin{tikzpicture}[scale=0.6]
    \begin{axis}[
        ybar,
        ymin=0, ymax=25,
        height=7.5cm,
        width=10cm,
        ytick={ 1,2,4,6,8,10,12,14,22},
      x tick label style={
      rotate=60,anchor=east,font=\scriptsize},
        symbolic x coords={
           USA-CTR, USA-FULL, Orkut,Sinaweibo,Rand-75M,Rand-100M,Rand-125M, Rmat-40M,Rmat-50M
            },
        xtick=data,
        bar width=0.1cm,
        xticklabel style={rotate=1},
        legend style={
            at={(0.2,0.7)},
            anchor=south,
            legend columns=1,
           /tikz/every even column/.append style={font=\tiny}
        },
        legend entries={Galois, Totem, Falcon, Green-Marl},
        legend image code/.code={%
        \draw[#1] (0cm,-0.1cm) rectangle (0.1cm,0.1cm);
      },  
      every node near coord/.append style={color=black, rotate=90, anchor=south, font=\tiny},
             every axis/.append style={font=\small},
    ]
    \addplot [ fill=yellow] coordinates {
      (USA-CTR,6.43)
      (USA-FULL,6.97) 
      (Orkut, 12.46)
      (Sinaweibo,9.78)
      (Rand-75M,10.35) 
      (Rand-100M,10.32) 
      (Rand-125M,10.45) 
      (Rmat-40M,9.81) 
      (Rmat-50M,9.49) 
            };
    \addplot [ fill=black] coordinates {
      (USA-CTR,1.09)
      (USA-FULL,0.82) 
      (Orkut, 7.54)
      (Sinaweibo,1.69)
      (Rand-75M,18.28) 
      (Rand-100M,18.41) 
      (Rand-125M,18.5) 
      (Rmat-40M,12.2) 
      (Rmat-50M,13.95) 
            };
    \addplot [ fill=red] coordinates {
      (USA-CTR,2.17)
      (USA-FULL,3.25) 
      (Orkut, 4.61)
      (Sinaweibo,6.07)
      (Rand-75M,7.47) 
      (Rand-100M,7.69) 
      (Rand-125M,8.01) 
      (Rmat-40M,7.02) 
      (Rmat-50M,7.09) 
            };
    \addplot [ fill=blue] coordinates {
      (USA-CTR,1.15)
      (USA-FULL,0.66) 
      (Orkut, 12.35)
      (Sinaweibo,8.1)
      (Rand-75M,7.06) 
      (Rand-100M,6.79) 
      (Rand-125M,6.66) 
      (Rmat-40M,8.17) 
      (Rmat-50M,7.76) 
            };
    \legend{\large{Galois},\large{Totem}, \large{Falcon},\large{Green-Marl}}
    \end{axis}
  \end{tikzpicture}
  \caption{BFS Speedup Over Galois single thread in CPU}
  \label{expt:bfscpucompare1}
\end{subfigure}%
\begin{subfigure}{0.5\linewidth}
\centering
  \begin{tikzpicture}[scale=0.7]
    \begin{axis}[
         ybar,
        ymin=0.01, ymax=28,
        height=6.0cm,
        width=9cm,
        ymode=log,
        log origin=infty,
        log basis y={2},
        x tick label style={
        rotate=60,anchor=east,font=\scriptsize},
        symbolic x coords={
           USA-CTR, USA-FULL, Orkut,Sinaweibo, Rand-75M,Rand-100M,Rand-125M, Rmat-30M,Rmat-40M,Rmat-50M
            },
        xtick=data,
        ytick={0,0.25, 0.5, 1, 2,4,8,16},
        bar width=0.1cm,
        xticklabel style={rotate=1},
        legend style={
            at={(0.8,0.77)},
            anchor=south,
            legend columns=1,
           /tikz/every even column/.append style={font=\tiny}
        },
        legend entries={Lonestar-GPU,Falcon,Gunrock},
        legend image code/.code={%
      \draw[#1] (0cm,-0.1cm) rectangle (0.1cm,0.1cm);
    },  
        every node near coord/.append style={color=black, rotate=90, anchor=south, font=\tiny},
             every axis/.append style={font=\small},
    ]
    \addplot [ fill=yellow] coordinates {
      (USA-CTR,0.14)
      (USA-FULL,0.19) 
      (Orkut, 0.05)
      (Sinaweibo,0.13)
      (Rand-75M,0.001) 
      (Rand-100M,0.001) 
      (Rand-125M,0.001) 
      (Rmat-40M,0.19) 
      (Rmat-50M,0.16) 
            };
    \addplot [ fill=black] coordinates {
      (USA-CTR,1.57)
      (USA-FULL,1.29) 
      (Orkut, 0.40)
      (Sinaweibo,1.64)
      (Rand-75M,2.09) 
      (Rand-100M,1.98) 
      (Rand-125M,1.89) 
      (Rmat-40M,1.06) 
      (Rmat-50M,1.01) 
            };
    \addplot [ fill=red] coordinates {
      (USA-CTR,7.89)
      (USA-FULL,11.39) 
      (Orkut, 0.27)
      (Sinaweibo,0.88)
      (Rand-75M,1.02) 
      (Rand-100M,0.001) 
      (Rand-125M,0.001) 
      (Rmat-40M,0.81) 
      (Rmat-50M,0.78) 
            };
    \legend{\large{Lonestar-GPU}, \large{Falcon}, \large{Gunrock}}
    \end{axis}
  \end{tikzpicture}
  \label{expt:bfsgpucompare-log}
  \caption{BFS Speedup over Totem in GPU}
\end{subfigure}
\caption{BFS comparison}
\label{expt:bfscpugpu}
\end{figure*}
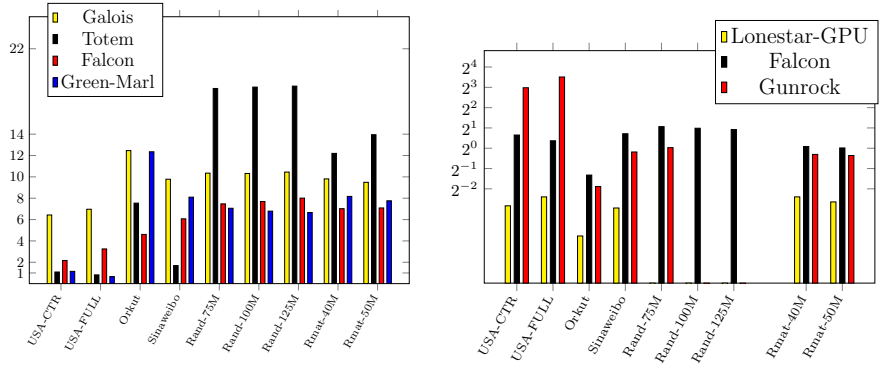

\begin{figure*}[h]
  \begin{subfigure}{0.5\textwidth}
   \begin{tikzpicture}[scale=0.8]
    \begin{axis}[
         ybar,
        ymin=0, ymax=20,
        height=5.5cm,
        width=8cm,
ytick={ 1,2,4,6,8,10,12,18},
x tick label style={
rotate=60,anchor=east,font=\scriptsize},
        symbolic x coords={
           USA-CTR, USA-FULL, Orkut,Sinaweibo,Rand-75M,Rand-100M,Rand-125M,Rmat-40M,Rmat-50M
            },
       xtick=data,
        bar width=0.1cm,
         xticklabel style={rotate=1},
       legend style={
            at={(0.8,0.7)},
            anchor=south,
            legend columns=1,
           /tikz/every even column/.append style={font=\tiny}
        },
        legend entries={Lonestar-GPU,Falcon,Gunrock},
        legend image code/.code={%
      \draw[#1] (0cm,-0.1cm) rectangle (0.1cm,0.1cm);
   },  
 every node near coord/.append style={color=black, rotate=90, anchor=south, font=\tiny},
             every axis/.append style={font=\small},
    ]
    \addplot [ fill=yellow] coordinates {
      (USA-CTR,6.67)
      (USA-FULL,5.86) 
      (Orkut, 12.01)
      (Sinaweibo,4.47)
      (Rand-75M,6.82) 
      (Rand-100M,7.34) 
      (Rand-125M,7.62) 
      (Rmat-40M,7.55) 
      (Rmat-50M,7.71) 
            };
    \addplot [ fill=black] coordinates {
      (USA-CTR,3.18)
      (USA-FULL,3.18) 
      (Orkut, 8.17)
      (Sinaweibo,4.59)
      (Rand-75M,3.3) 
      (Rand-100M,3.8) 
      (Rand-125M,3.94) 
      (Rmat-40M,5.78) 
      (Rmat-50M,6.17) 
            };
    \addplot [ fill=red] coordinates {
      (USA-CTR,0.13)
      (USA-FULL,0.08) 
      (Orkut, 17.33)
      (Sinaweibo,1.55)
      (Rand-75M,7.63) 
      (Rand-100M,7.26) 
      (Rand-125M,7.4) 
      (Rmat-40M,5.96) 
      (Rmat-50M,5.69) 
            };
 \legend{\large{Galois},\large{Totem}, \large{Falcon}}
 \end{axis}
   \end{tikzpicture}
\caption{CPU: Speedup over Galois single thread}
\label{expt:cccpucompare}
\end{subfigure}%
  \begin{subfigure}{0.5\textwidth}
\centering
   \begin{tikzpicture}[scale=0.8]
    \begin{axis}[
         ybar,
        ymin=0.01, ymax=180,
        height=5.5cm,
        width=8cm,
ymode=log,
log origin=infty,
        log basis y={2},
x tick label style={
rotate=60,anchor=east,font=\scriptsize},
        symbolic x coords={
           USA-CTR, USA-FULL, Orkut,Sinaweibo,Rand-75M,Rand-100M,Rand-125M,Rmat-40M,Rmat-50M
            },
       xtick=data,
ytick={0,1,2,4,8,16,150},
        bar width=0.1cm,
         xticklabel style={rotate=1},
       legend style={
            at={(0.8,0.77)},
            anchor=south,
            legend columns=1,
           /tikz/every even column/.append style={font=\tiny}
        },
        legend entries={Lonestar-GPU,Falcon,Gunrock},
        legend image code/.code={%
      \draw[#1] (0cm,-0.1cm) rectangle (0.1cm,0.1cm);
   },  
 every node near coord/.append style={color=black, rotate=90, anchor=south, font=\tiny},
             every axis/.append style={font=\small},
    ]
    \addplot [ fill=black] coordinates {
      (USA-CTR,4.74)
      (USA-FULL,4.06) 
      (Orkut, 0.62)
      (Sinaweibo,1.61)
      (Rand-75M,2.48) 
      (Rand-100M,2.07) 
      (Rand-125M,1.82) 
      (Rmat-40M,0.76) 
      (Rmat-50M,0.72) 
            };
    \addplot [ fill=red] coordinates {
      (USA-CTR,84)
      (USA-FULL,142) 
      (Orkut, 0.78)
      (Sinaweibo,1.44)
      (Rand-75M,3.33) 
      (Rand-100M,2.96) 
      (Rand-125M,2.63) 
      (Rmat-40M,0.48) 
      (Rmat-50M,0.49) 
            };
 \legend{Falcon, Gunrock}
 \end{axis}
   \end{tikzpicture}
\caption{GPU: Speedup over Totem}
\label{expt:bfsgpucompare-log2}
\end{subfigure}
\caption{CC comparison}
\label{expt:cccpugpu}
\vspace{0.2in}
  \end{figure*}
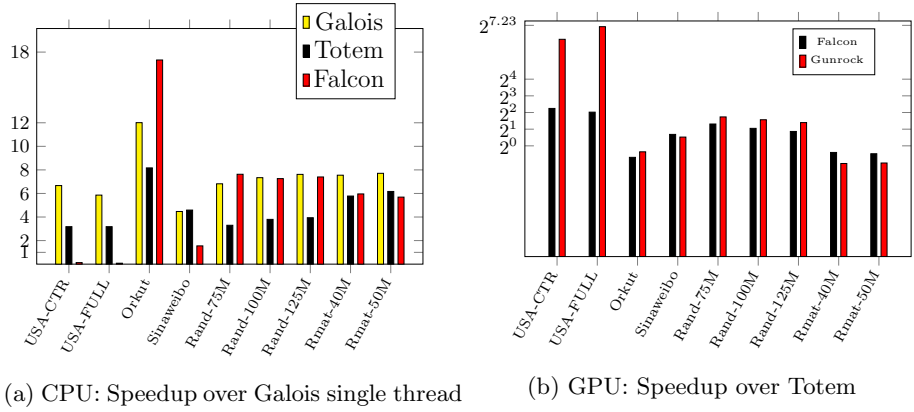

Figures~\ref{expt:ssspcpugpu}, \ref{expt:bfscpugpu}, \ref{expt:cccpugpu} present the performance benefit of our modified Falcon against other frameworks.
The GPU-baseline used for this comparison is Totem, whose speedup is assumed to be 1.0 (hence not shown in the plots).
On CPU, the baseline is Galois with one thread.
For GPU-SSSP, we observe that Falcon-generated code provides consistently better speedups compared to LonestarGPU and Gunrock, except on the two social networks (orkut and sinaweibo).
Totem performs better on the social networks as well as on RMAT graphs due to its inbuilt edge-based processing and other optimizations to improve load-balancing across GPU threads.
For BFS, the results are mixed across various frameworks and there is no clear winner, but there are interesting patterns based on the graph types.
Gunrock performs quite well on the road networks (USA-full and USA-CTR), primarily due to its work-efficient worklist-based processing.
Totem outperforms again on social networks due to edge-based processing and better load-balancing.
Performance of almost all the frameworks on RMAT graphs is quite similar, with LonestarGPU performing poorly.
Our Falcon stands out on random graphs with speedups close to 2$\times$ over all other frameworks.
On CPU, Galois outperforms other frameworks for SSSP, but Green-Marl is a close second.
Note that Galois uses hand-crafted libraries, while Green-Marl is auto-generated, similar to Falcon.
Totem performs quite poorly for SSSP, but bounces back for BFS outperforming all the other frameworks.
Galois outperforms other frameworks on road networks because it uses a different algorithm (delta-stepping).
For CC,  Falcon does not perform  well on road inputs. Performance of MST on GPU is shown in Figure~\ref{expt:mstcpucompare}.

\vspace{-0.1in}
\subsection{Effect of Vertex-based versus Edge-based}\label{expt:vertexedge}
\vspace{-0.1in}
Figures~\ref{expt:falconselfcpu} and ~\ref{expt:falconselfgpu} presents results of edge-based versus vertex-based processing of Falcon across various graphs for CC, BFS and SSSP. 
We observe that edge-based processing performs better in social-networks (orkut and sinaweibo) and RMAT graphs. 
Both these kinds of graphs have skewed (power-law) degree-distribution resulting in large load-imbalance with vertex-based processing.
These graphs follow small-world property due to this peculiar (and natural) degree distribution.
On GPUs, this load-imbalance manifests itself as thread-divergence as the number of iterations (based on the number of neighbors) of each thread has high variance.
In other words, threads mapped to vertices having few neighbors have to wait for others mapped to high-degree vertices. 
This inhibits parallelism for SIMT style of processing. 
In contrast, in edge-based processing, since threads are mapped to (a group of) edges, the load-imbalance is relatively negligible.
This results in better thread-divergence among warp-threads, leading to improved execution time.
Road networks and random graphs, on the other hand, have quite uniform degree-distribution.
Therefore, edge-based processing is not very helpful.
In fact, for uniform degree-distributions, edge-based processing may lead to inferior results (as seen in our experiments), due to increased synchronization requirement. 
Different outgoing edges of a vertex are processed sequentially by the same thread in vertex-based processing; whereas, those are processed in parallel by different threads.
Thus, edge-based processing necessitates more coordination among threads with respect to reading and updating attribute values of vertices. The worklist based code does not benefit on GPU. 
Speedup of  $\Delta$-Stepping worklist~\cite{Meyer:1998:DPS:647908.740136} based code is much faster than    {\it OpenMP} library based vertex-based code as shown in figure~\ref{expt:worklist-cpu}.
\input{newfig.tex}
We observe that, unlike on GPUs, edge-based processing is not helpful on CPUs. 
This is primarily due to CPUs not having enough resources to utilize the additional parallelism exposed by edge-based processing.
Thus, a few tens of threads perform in a similar manner in the presence of a million vertices or multi-million edges.
The only exception is sinaweibo graph which witnesses over $3\times$ speedup for CC on CPU due to edge-based processing.
The improvements on this graph are also high for other algorithms as well (BFS and SSSP) compared to other graphs.
This occurs due to higher average degree in this social network.
Higher average degree adds sequentiality in vertex-based processing, while edge-based processing is not amenable to degree-distribution or average degree.
The overall effect gets pronounced for such dense graphs.
\vspace{-0.2in}
\subsection{Effect of Synchronous versus Asynchronous Processing}\label{expt:syncasync}
\vspace{-0.1in}
\begin{minipage}{0.5\linewidth}
\begin{table}[H]
\centering
\footnotesize
\begin{tabular}{|l|r|r|}
\hline
\textbf{Graph} & \textbf{\shortstack {Synch-\\ronous}}  & \textbf{\shortstack{Asyn-\\chronous}}\\
\hline
USA-CTR &	  86	& 91 	\\
USA-full &	  134	& 106   \\
orkut &   425	& 285   \\
sinaweibo &   8310	& 4508  \\
rand-25M &   	  458	& 347   \\
rand-50M &   	  945	& 741   \\
rmat-10M &   	  329	& 320   \\
rmat-20M &   	  771	& 665   \\
\hline
\textbf{Average}	&	1433	& 882	\\
\hline
\end{tabular}
\caption{BFS and SSSP (CPU)  Sync. versus Async.}
\label{tab:async-cpu}
\end{table}
\end{minipage}%
\begin{minipage}{0.5\linewidth}
    \begin{tikzpicture}[scale=0.7]
      \begin{axis}[
          ybar,
          ymin=0, ymax=20,
          height=6.0cm,
          width=9cm,
          ytick={ 1,2,4,6,8,10,12,14,16,18},
          x tick label style={
          rotate=60,anchor=east,font=\scriptsize},
          symbolic x coords={
            USA-CTR, USA-FULL, Orkut,Sinaweibo, Rand-25M,Rand-50M,Rand-75M,Rand-100M,Rand-125M, Rmat-10M,Rmat-20M,Rmat-30M,Rmat-40M,Rmat-50M
          },
          xtick=data,
          bar width=0.15cm,
          xticklabel style={rotate=1},
          legend style={
            at={(0.8,0.7)},
            anchor=south,
            legend columns=1,
            /tikz/every even column/.append style={font=\large}
          },
          legend entries={Galois, Falcon},
          legend image code/.code={%
            \draw[#1] (0cm,-0.1cm) rectangle (0.1cm,0.1cm);
          },  
          every node near coord/.append style={color=black, rotate=90, anchor=south, font=\large},
          every axis/.append style={font=\large},
        ]
        \addplot [ fill=yellow] coordinates {
          (USA-CTR,8.44)
          (USA-FULL,8.88) 
          (Orkut, 7.09)
          (Sinaweibo,10.67)
          (Rand-75M,10.24) 
          (Rand-100M,12.03) 
          (Rand-125M,10.11) 
          (Rmat-40M,8.47) 
          (Rmat-50M,8.12) 
        };

        \addplot [ fill=black] coordinates {
          (USA-CTR,10.30)
          (USA-FULL,10.78) 
          (Orkut, 17.45)
          (Sinaweibo,10.17)
          (Rand-75M,6.18) 
          (Rand-100M,7.5) 
          (Rand-125M,6.16) 
          (Rmat-40M,0.75) 
          (Rmat-50M,1.57) 
        };

        \legend{\large{Galois}, \large{Falcon}}
      \end{axis}
    \end{tikzpicture}
\captionof{figure}{MST Speedup over Galois Single}
   \label{expt:mstcpucompare}
\end{minipage}

Table~\ref{tab:async-cpu} presents the effect of asynchronous processing for various graphs on CPU.
We used a combination of BFS and SSSP to perform independent processing on the same graph.
We observe that asynchronous version improves execution time by 38\%. 
This occurs because threads do not have to wait for other threads.
This is primarily true on CPUs as threads are monolithically working on different parts of the graph and seldom require synchronization.
In our experiments, since all the GPU resources were utilized by a single kernel, asynchronous processing performed similar to synchronous exeuction.

\vspace{-0.1in}
\vspace{-0.1in}
\subsection{Effect of Code Generation for Multiple Targets}\label{expt:cpugpu}
\vspace{-0.1in}
\begin{table}
\parbox{0.45\linewidth}{
\centering
\footnotesize
\begin{tabular}{|l|r|r|r|}
\hline
\textbf{Graphs} & \textbf{CPU-16} & \textbf{GPU} & \textbf{\shortstack{multi-\\GPU}}\\\hline
\shortstack{rand-25M +\\rand50M} & 3866 & 1234 & 826 \\
\hline
\shortstack{rmat-10M +\\rmat20M} & 3024 & 1176 & 792 \\
\hline
\end{tabular}
\caption{Connected Components}
\label{tab:multigpucc}
}
\hfill
\parbox{0.45\linewidth}{
\centering
\footnotesize
\begin{tabular}{|l|r|r|r|}
\hline
\textbf{Graphs} & \textbf{CPU-16} & \textbf{GPU} & \textbf{\shortstack{multi-\\GPU}}\\\hline
\shortstack{USA-CTR +\\USA-full} & 25363 & 12569 & 9138 \\
\hline
\shortstack{sinaweibo +\\orkut} & 4845 & 1503 & 1259 \\
\hline
\end{tabular}
\caption{Breadth First Search}
\label{tab:multigpubfs}
}
\captionof{figure}{Execution time (in ms) of BFS and CC for various targets}
\end{table}
\vspace{-0.1in}
 Our approach can seamlessly generate code for CPU or GPU or multi-GPU.
The multi-GPU code works with different graphs for the same algorithm, or with the same graph for different algorithms.
Table~\ref{tab:multigpucc} presents results for the former with CC as the algorithm generating code for CPU with 16 threads, single GPU and two GPUs, while Table~\ref{tab:multigpubfs} presents those for BFS.
We observe that multi-GPU version took much less time as compared to other backends. 
In both CPU and single-GPU versions, the graphs are processed one after another.
On the other hand, in multi-GPU version, both the graphs are processed simultaneously in different GPUs; so the overall execution time is the larger of the two.

\vspace{-0.1in}

%% file: newfig.tex
\begin{figure*}
\centering
\begin{subfigure}{0.25\linewidth}
  \begin{tikzpicture}[scale=0.7]
    \begin{axis}[
        ybar,
        ymin=0.001, ymax=4,
        height=5.5cm,
        width=5cm,
        ytick={ 0,0.5,1,2,3,4},
        x tick label style={
        rotate=60,anchor=east,font=\scriptsize},
        symbolic x coords={
  Orkut,Sinaweibo,Rmat-40M,Rmat-50M
            },
        xtick=data,
        bar width=0.1cm,
        xticklabel style={rotate=1},
        legend style={
            at={(0.9,0.60)},
            anchor=south,
            legend columns=1,
           /tikz/every even column/.append style={font=\tiny}
        },
        legend entries={Lonestar-GPU,Falcon,Gunrock},
        legend image code/.code={%
      \draw[#1] (0cm,-0.1cm) rectangle (0.1cm,0.1cm);
    },  
        every node near coord/.append style={color=black, rotate=90, anchor=south, font=\normalsize},
            every axis/.append style={font=\small},
    ]
    \addplot [ fill=yellow] coordinates {
      (Orkut, 0.38)
      (Sinaweibo,1.17)
      (Rmat-40M,0.46) 
      (Rmat-50M,0.45) 
            };
    \addplot [ fill=black] coordinates {
      (Orkut, 1.04)
      (Sinaweibo,3.16)
      (Rmat-40M,0.94) 
      (Rmat-50M,0.99) 
            };
    \addplot [ fill=red] coordinates {
      (Orkut, 0.53)
      (Sinaweibo,1.9)
      (Rmat-40M,0.79) 
      (Rmat-50M,0.72) 
            };
    \legend{\large{bfs}, \large{cc}, \large{sssp}}
    \end{axis}
  \end{tikzpicture}
  \caption{ edge-based-CPU}
  \label{expt:falconselfcpu}
\end{subfigure}%
\begin{subfigure}{0.25\linewidth}
\centering
  \begin{tikzpicture}[scale=0.7]
      \begin{axis}[
          ybar,
          ymin=0.001, ymax=4,
          height=5.5cm,
          width=5cm,
          ytick={ 0,0.5,1,2,3,4},
          x tick label style={
          rotate=60,anchor=east,font=\scriptsize},
            symbolic x coords={
               Orkut,Sinaweibo,Rmat-40M,Rmat-50M
                },
          xtick=data,
          bar width=0.1cm,
          xticklabel style={rotate=1},
          legend style={
                at={(0.9,0.6)},
                anchor=south,
                legend columns=1,
               /tikz/every even column/.append style={font=\tiny}
            },
            legend entries={BFS,CC, SSSP},
            legend image code/.code={%
          \draw[#1] (0cm,-0.1cm) rectangle (0.1cm,0.1cm);
      },  
      every node near coord/.append style={color=black, rotate=90, anchor=south, font=\normalsize},
                 every axis/.append style={font=\small},
    ]
        \addplot [ fill=yellow] coordinates {
          (Orkut, 1.47)
          (Sinaweibo,2.64)
          (Rmat-40M,1.12) 
          (Rmat-50M,1.18) 
                };
        \addplot [ fill=black] coordinates {
          (Orkut, 2.6)
          (Sinaweibo,2.46)
          (Rmat-40M,1.46) 
          (Rmat-50M,1.60) 
                };
        \addplot [ fill=red] coordinates {
          (Orkut, 1.41)
          (Sinaweibo,2.13)
          (Rmat-40M,1.15) 
          (Rmat-50M,1.24) 
                };
     \legend{\large{bfs}, \large{cc}, \large{sssp}}
    \end{axis}
  \end{tikzpicture}
  \caption{edge-based-GPU}
  \label{expt:falconselfgpu}
\end{subfigure}
\begin{subfigure}{0.3\linewidth}
\centering
  \begin{tikzpicture}[scale=0.7]
    \begin{axis}[
        ybar,
        ymin=0.01, ymax=400,
        height=5.5cm,
        width=5cm,
        ytick={ 1,2,4,6,8,10,100,200},
        ymode=log,
        log origin=infty,
        log basis y={2},
        x tick label style={
        rotate=60,anchor=east,font=\scriptsize},
        symbolic x coords={
          USA-CTR, USA-FULL, Orkut,Sinaweibo
        },
        xtick=data,
        bar width=0.1cm,
        xticklabel style={rotate=1},
        legend style={
            at={(0.8,0.7)},
            anchor=south,
            legend columns=1,
           /tikz/every even column/.append style={font=\tiny}
        },
        legend entries={BFS, CC, SSSP},
        legend image code/.code={%
      \draw[#1] (0cm,-0.1cm) rectangle (0.1cm,0.1cm);
    },
        every node near coord/.append style={color=black, rotate=90, anchor=south, font=\tiny},
             every axis/.append style={font=\small},
    ]

    \addplot [ fill=yellow] coordinates {
      (USA-CTR,12.69)
      (USA-FULL,27.45) 
      (Orkut, 0.59)
      (Sinaweibo,2.01)
            };
    \addplot [ fill=black] coordinates {
      (USA-CTR,123.19)
      (USA-FULL,181.18) 
      (Orkut, 1.22)
      (Sinaweibo,6.90)
            };
    \addplot [ fill=red] coordinates {
      (USA-CTR,81.50)
      (USA-FULL,181.18) 
      (Orkut, 0.62)
      (Sinaweibo,2.34)
            };
    \legend{\large{bfs}, \large{cc}, \large{sssp}}
    \end{axis}
  \end{tikzpicture}
  \caption{worklist-based-CPU}
  \label{expt:worklist-cpu}
\end{subfigure}
\caption{Speedup of edge-based  and worklist-based code over vertex-based code}
  \label{expt:edgevertex}
\vspace{0.1in}
\end{figure*}

%% file: related.tex
\vspace{-0.1in}
\section{Related Work}\label{sec:related}
\vspace{-0.05in}
\par Green-Marl~\cite{Hong:2012:GDE:2150976.2151013} is a graph DSL for 
implementing parallel graph algorithms on multi-core CPUs. 
Green-Marl was extended for GPUs~\cite{lighthouse} and CPU clusters~\cite{Hong:2014:SSG:2581122.2544162}.
 Falcon~\cite{falcon} is a DSL for graph analytics on single machine with one or more GPUs. Falcon is extended for distributed systems with CPU and GPU~\cite{dhfalcon},~\cite{heteropar}.  
Galois~\cite{Pingali:2011:TPA:1993316.1993501} is a C++ framework for  graph analytics on multi-core CPUs. 
Ligra~\cite{Shun:2013:LLG:2517327.2442530} is a framework for writing graph traversal algorithms
for multi-core shared memory systems.
 \par 
There are efficient  implementations different graph algorithms ~\cite{Merrill:2012:SGG:2370036.2145832,sariyuce13,mendezlojo12}  on \GPU.
Worklist-based graph algorithms  do 
not benefit much on \GPU~\cite{DAVIDSON2014}.
The Lonestar-GPU~\cite{nasre13:MAG:2517327.2442531} framework supports  dynamic graph algorithms on \GPU. 
The Gunrock~\cite{Wang:2016:GHG:3016078.2851145} framework provides  a data-centric abstraction for graph operations 
 with   GPU-specific optimizations.
Totem~\cite{Gharaibeh:2012:YOT:2370816.2370866} is a heterogeneous framework for graph processing for a multi-GPU machine.
\par
Large graphs require processing on computer cluster.
GraphLab~\cite{Low:2012:DGF:2212351.2212354}, PowerGraph~\cite{Gonzalez:2012:PDG:2387880.2387883}, Pregel~\cite{Malewicz:2010:PSL:1807167.1807184} and  Giraph~\cite{Ching:2015:OTE:2824032.2824077} are popular distributed graph analytics framework.
 Bulk Synchronous Parallel (BSP) Model  \cite{Valiant:1990:BMP:79173.79181} of execution and asynchronous executions are popular models of executions.
Gluon~\cite{Dathathri:2018:GCS:3192366.3192404} uses Galois and Ligra and generates distributed-memory versions of these systems.

%% file: conclusion.tex
\section{Conclusion}\label{sec:conclusion}
Irregular codes have data-dependent access patterns.
Therefore, compilers need to make pessimistic assumptions leading to very conservative code.
While DSLs for irregular codes allow us the flexibility to make more informed decisions about the domain, existing DSLs lack adaptability.
Different graphs expect different kinds of processing to achieve the best performance.
While existing DSLs do allow changing the algorithm specification to suit a purpose, it would be ideal if the specification remains intact and the compiler judiciously generates the necessary efficient code.
We presented our experiences in achieving the same, for a graph DSL, Falcon.
In particular, we auto-generated codes for vertex-based and edge-based, for synchronous versus asynchronous, for worklist-driven versus topology-driven, and for CPU versus GPU versus multi-GPU processing.
We illustrated the effectiveness of our techniques using a variety of algorithms and several real-world graphs.